# THE EFFECT OF ANOMALOUS ADSORPTION OF $H_2O$ AND $CO_2$ BY PRE-HYDRATED $YBa_2Cu_3O_{6.53}$

*Fetisov A. V.*[1], *Kozhina G.A.*[2], *Fetisov V.B.*[2], *Estemirova S.Kh.*[1], *Fedorova O.M.*[1], *Gulyaeva R.I.*[1]

[1]*Institute of Metallurgy, Ural Division RAS, Ekaterinburg, Russia*
[2]*Ural State University of Economy, Ekaterinburg, Russia*

**Abstract**

The results of a comprehensive study on the $YBa_2Cu_3O_{6.53}$ oxide subjected to "mild" hydration (exposure at small values of $p_{H_2O}$) by a special technique have been reported. The material modified in this way acquires an interesting property; namely, in natural conditions it intensively absorbs large quantities of components ($H_2O$ and $CO_2$) comprising the gas atmosphere. Arguments have been adduced that at least the water enters the crystal lattice of the yttrium–barium cuprate. This lattice is in the two-phase (tetragonal and orthorhombic) state: testing of the magnetic properties of the oxide has revealed the presence of antiferromagnetic and diamagnetic regions in this material.

## Introduction

The crystal structure of the high-temperature $YBa_2Cu_3O_{6+\delta}$ superconductor has an incomplete plane oxygen sublattice, which tends to ordering and the exchange of oxygen with the gaseous phase. Because of this feature, the yttrium–barium cuprate has been used for some time as a "model object" for investigations of the two-dimensional oxygen diffusion in oxides [1]. Numerous diffusion investigations mostly are in good agreement. According to these investigations, the coefficient for the chemical diffusion of oxygen in $YBa_2Cu_3O_{6+\delta}$ at 400°C is ~$10^{-12}$–$10^{-11}$ cm$^2$/s. However, in some studies, the oxygen exchange in the yttrium–barium cuprate was considerably intensified by exposing its surface to a strong liquid-phase reductant [2], laser radiation [3], etc.

Atmospheric water can also enter the structural sites of the incomplete oxygen sublattice of $YBa_2Cu_3O_{6+\delta}$ mainly in the form of hydroxide–ions and protons [4]. In natural conditions this process is rather slow (usually, powders of this oxide are loaded with 1 mass % moisture in several months [5]). The superconductor properties of the yttrium–barium cuprate therewith change insignificantly [4, 5]. A synergetic effect – the adsorption of the carbon dioxide requires that water is adsorbed first – is observed when $YBa_2Cu_3O_{6+\delta}$ is exposed simultaneously to water vapor and $CO_2$ [6]. However, large molecules of the carbon dioxide practically do not enter the oxide structure. The oxide decomposition products are formed when more than 1.3 mass % moisture is dissolved in the structure of the yttrium–barium cuprate [4].

It will be shown below that preliminary "mild" hydration of $YBa_2Cu_3O_{6+\delta}$ by the method [7] endows the oxide with the ability to intensively adsorb atmospheric components, namely, $H_2O$ and $CO_2$. As this takes place, the yttrium–barium cuprate passes to the two-phase (tetragonal and orthorhombic) state with the magnetic properties characteristic of these phases already at the preliminary stage of hydration.

**Preparation of Samples**

The object of study was prepared from barium nitrate $Ba(NO_3)_2$ (chemically pure), yttrium oxide $Y_2O_3$ and basic cupric carbonate $CuCO_3 \times Cu(OH)_2$ (highly pure) by calcination at 525–725–925°C for 120 h. According to the X-ray diffraction analysis, the ready yttrium–barium cuprate contained ~95% of the main phase and the balance $Y_2BaCuO_5$. At the final stage of heat treatment at 700°C, the material was oxidized in a flow of pure air. According to the X-ray diffraction analysis, the parameters of a unit cell in the oxidized yttrium–barium cuprate were $a$ = 3.8442(10), $b$ = 3.8895(11), and $c$ =

11.7535(32) Å. The oxygen concentration of $YBa_2Cu_3O_{6+\delta}$ was evaluated at 6.53 from the change of the mass resulting from the long-time holding of the sample at 400°C on the assumption that in the final state the parameter $\delta \approx 0.93$.

The obtained material was dispersed to a size of ~20 μm and was modified by a special technology [7] at $t$ = 25°C, $p_{O_2}$ = 20 kPa, and $p_{H_2O}$ = 110±5 Pa, leading to the formation of a compound phase with "special" transport properties. An illustrative example of these properties can be an anomalously intensive exchange of oxygen between the yttrium–barium cuprate and the gaseous phase at 25°C [8] and 400°C [9].

**Experimental**

The modified yttrium–barium cuprate was saturated with moisture and $CO_2$ in standard conditions at $p_{H_2O}$ = 0.5÷2.3 kPa and $p_{CO_2}$ = 0.03 kPa. The kinetic curve of the process, which was plotted on the basis of the weighing data (a Shimadzu AUW 120D analytical basis), was a linear dependence showing the mass of the sample as a function of time, Fig. 1. When $p_{H_2O}$ = 2.3 kPa, the process stopped at degrees of saturation close to ($H_2O$+$CO_2$), 2 moles/ $YBa_2Cu_3O_{6+\delta}$, 1 mole. The limiting degree of saturation was reduced quickly with the elasticity of the water vapor. The effect of $p_{H_2O}$ on the rate of the gas absorption is shown in Fig. 2.

Specific features of desorption of the absorbed gases resulting from the reduction of $p_{H_2O}$ are shown in Fig. 3. The thermogravimetric, differential scanning calorimetric, and mass spectrometric curves (a Netzsch thermoanalytical setup comprised of an STA 449C Jupiter thermal analyzer and a QMS 403C Aëolos mass spectrometer) in Figs. 4 and 5 show degassing of the gas-saturated material during heat treatment.

The surface of the particles of gas-saturated $YBa_2Cu_3O_{6.53}$ was examined in an EVO 40 XVP scanning electron microscope (Carl Zeiss) with "direct" electrons. Since the SEM analysis was performed in a high vacuum (~$10^{-4}$Pa), the test sample first was heated for a short time for degassing. The images of two points on the oxide surface are given in Fig. 6.

As the sample was saturated with gases, its X-ray diffraction analysis was performed in an XRD 7000 diffractometer (Shimadzu Corp.) (Fig. 7 and Table). Continuous scanning with a step of 0.02 deg in $CuK_{\alpha}$ radiation was used. A silicon powder served as the external standard.

The low-temperature magnetic measurements (Figs. 8 and 9) were made using an automated Cryogenic CFS-9T-CVTI VSM system (UK).

**Discussion**

From a comparative analysis of the rates, at which the yttrium–barium cuprate was saturated with gases before and after its modification, it followed (see Fig. 1) that "special" transport properties of this oxide show up not only with respect to oxygen, but also to $H_2O$ and $CO_2$. A mass spectrometric analysis of the thermal desorption products of modified $YBa_2Cu_3O_{6.53}$ (Figs. 4 and 5) confirmed that the oxide contained a large quantity of water and carbon dioxide after short-time hydration.

The SEM images of the surface in Fig. 6 can provide proof that such a large quantity of the adsorbed gases is not consumed for saturation of the assumed microscopic pores of the material. Steps and microcracks can only be seen on the surface of coarse $YBa_2Cu_3O_{6.53}$ particles in these photographs. Furthermore, no phase contrast is observed, pointing to the phase homogeneity of the material surface exposed to corrosive $H_2O$ and $CO_2$ gases. At the same time, the X-ray diffraction data (see Fig. 7) reveal a low decomposition degree of the yttrium–barium cuprate throughout its saturation with water and carbon dioxide. The arrows in the fig. 7 mark lines related to $BaCO_3$ compound. Its amount evaluated for highly saturated sample is ~ 7 vol.%, that corresponds to 1.2 mass% of $CO_2$. On the basis of TGA data, precisely this quantity of carbonic acid gas is evolved while annealing the sample at 900°C. Apparently, phase homogeneity of the surface of $YBa_2Cu_3O_{6.53}$ presupposes nearly full saturation of the surface by barium carbonate. The SEM and the X-ray results suggest the absence of a large quantity of water on the surface of the oxide during its saturation; i.e., the water fills the crystal structure.

In turn, the intensity of the reflections in the X-ray diffraction patterns, Fig. 7, does not change as the $YBa_2Cu_3O_{6.53}$ structure is increasingly saturated with the gases. This should affect, first of all, (013), (110) and (103) lines, which reflect the occupancy of sites in the incomplete oxygen sublattice of the oxide. It seems likely that $H_2O$ particles do not occupy strictly determined positions in the structure of modified $YBa_2Cu_3O_{6.53}$, but form their own "amorphous medium" therein.

The X-ray diffraction results listed in the table do not indicate any structural phase changes proceed during "mild" pre-hydration of the oxide: it was produced material radiographically identical to the initial single phase orthorhombic structure.

The curves for absorption–desorption of the gases at different $p_{H_2O}$, Fig. 3, show the time dependence of the $H_2O$ and $CO_2$ state in the yttrium–barium cuprate. The state changes towards stronger binding with the lattice. A similar conclusion follows from the thermal desorption data in Figs. 4 and 5 since the maximum temperatures for the release of water and carbon dioxide from $YBa_2Cu_3O_{6.53}$ increase with the time of their stay in the oxide structure.

From the TG curves in Figs. 4 and 5 it is possible to evaluate the mass ratio of $H_2O$ and $CO_2$ at different stages of their adsorption in the oxide. It turns out that this ratio is 65±3% $H_2O$ and 35±3% $CO_2$ regardless of the stage. An analogous result follows from an analysis of the TG curves of the samples saturated with the gases at smaller $p_{H_2O}$ (the curves are not shown). Thus, the lower the elasticity of the water vapor, the less $H_2O$ and $CO_2$ are equally adsorbed. This regularity points to the synergetic effect in the modified yttrium–barium cuprate: $CO_2$ diffuses in the oxide only in combination with $H_2O$. In this connection it should be noted that an exponential growth of the adsorption rate of the gases with increasing $p_{H_2O}$ (see Fig. 2) can be indicative of some field effect produced (probably, via the dipole electrical moment of its molecules) by water, which is adsorbed on the oxide surface, on the potential barriers for chemisorption of $CO_2$ and entering of $H_2O$ into the crystal structure.

Finally, the magnetic properties of the yttrium–barium cuprate, either modified or unmodified by preliminary "mild" hydration (see Figs. 8 and 9), were measured. It turned out that, as distinct from the "standard" sample, the magnetic moment of the modified sample did not drop to negative values, which are characteristic of the superconducting state, below some critical temperature. Oppositely, a monotonic growth of M(T), which is typical of antiferromagnets (AFM), was observed. It is known [9] that this behavior is typical of the tetragonal structure of $YBa_2Cu_3O_{6+\delta}$ and is atypical of the orthorhombic structure. But since the X-ray diffraction analysis revealed namely orthorhombic state of the modified oxide under study, it was interesting to check if the AFM phase coexisted with the diamagnetic phase, which has an opposite, but a weaker, magnetic moment. To detect this phase, the external magnetic field was enhanced assuming that at some B the AFM moment should and the diamagnetic moment should not flatten out. Figure 9 presents the dependence M(T) measured for the modified yttrium–barium cuprate in a field of 1 T. It is seen that the curve exhibiting the AFM behavior ends with a diamagnetic "tail", which confirms that the modified $YBa_2Cu_3O_{6+\delta}$ orthorhombic phase possesses superconductivity.

To our mind, the most important result of the magnetic measurements is an indication the two structural phases (tetragonal and orthorhombic, hardly distinguishable by radiographic means in our case) coexist in the modified

yttrium–barium cuprate. This situation is similar to the spinodal decomposition [8] of the orthorhombic phase having an intermediate oxygen concentration into the orthorhombic and tetragonal phases, which can hardly be discerned by conventional X-ray diffraction method.

## Conclusions

Thus, it was shown that the yttrium–barium cuprate, which has been pre-hydrated by a special method, can intensively absorb large quantities of atmospheric components, namely, $H_2O$ and $CO_2$. The last substances strengthen their binding with the structure as they are staying in the oxide. The crystal lattice of pre-hydrated $YBa_2Cu_3O_{6.53}$ is in the two-phase (tetragonal and orthorhombic) state. Therewith, the oxide shows antiferromagnetic and diamagnetic properties simultaneously.

## Acknowledgments

We are grateful to A.B. Shubin, who took images of the $YBa_2Cu_3O_{6.53}$ surface in a scanning electron microscope, and G.A. Dorogina and V.Ya. Mitrofanov, who plotted the M(T) dependences and participated in their discussion.

This work was supported by the Russian Foundation for Basic Research (grant no. 07-03-00280). It was performed on the equipment at the Center of Collaborative Access "Ural–M".

*Table*

Lattice parameters of a gas-saturated $YBa_2Cu_3O_{6.53}$ sample pre-hydrated in "mild" conditions

| Gas saturation, mass % | Phase composition | Unit cell parameters, Å | | | $V$, Å$^3$ |
|---|---|---|---|---|---|
| | | *a* | *b* | *c* | |
| Initial, unhydrated material | O | 3.8399(8) | 3.8793(6) | 11.7324(16) | 174.769 (126) |
| 0 | O | 3.8376 (8) | 3.8789 (12) | 11.7299 (35) | 174.583 (153) |
| 0.46 | O | 3.8378 (8) | 3.8796 (9) | 11.7331 (26) | 174.692 (114) |
| 2.07 | O | 3.8374 (7) | 3.8786 (8) | 11.7275 (23) | 174.551 (103) |
| 2.91 | O | 3.8378 (10) | 3.8790 (12) | 11.7352 (35) | 174.699 (152) |

O: orthorhombic phase, space group Pmmm.

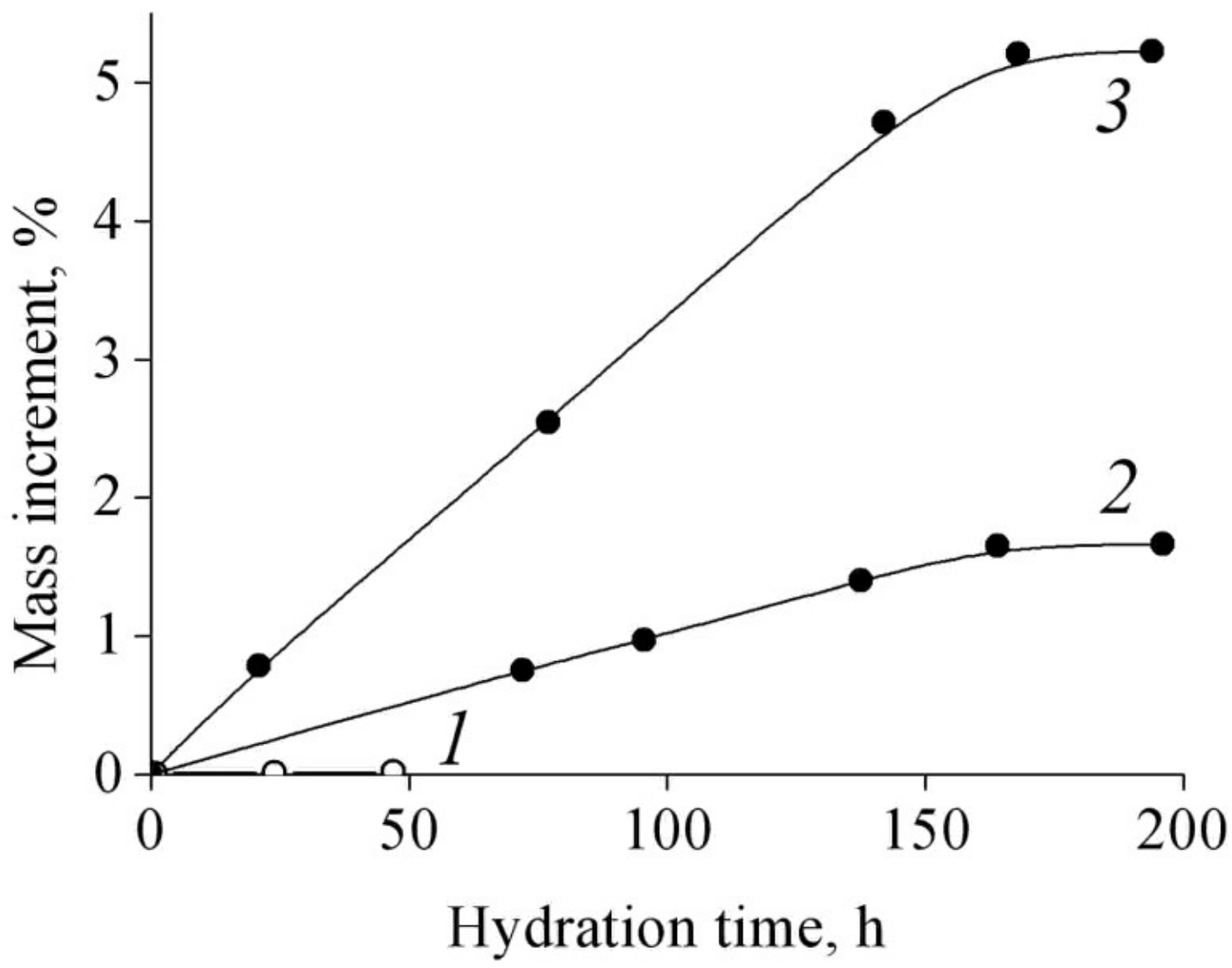

**Fig. 1.** Gas absorption in "standard" samples of $YBa_2Cu_3O_{6.53}$ (curve 1) and its samples modified as described in [7] (curves 2 and 3). The dependences *1* and *3* were obtained at $p_{H_2O}$ = 2.3 kPa; the dependence *2*, at $p_{H_2O}$ = 2.0 kPa.

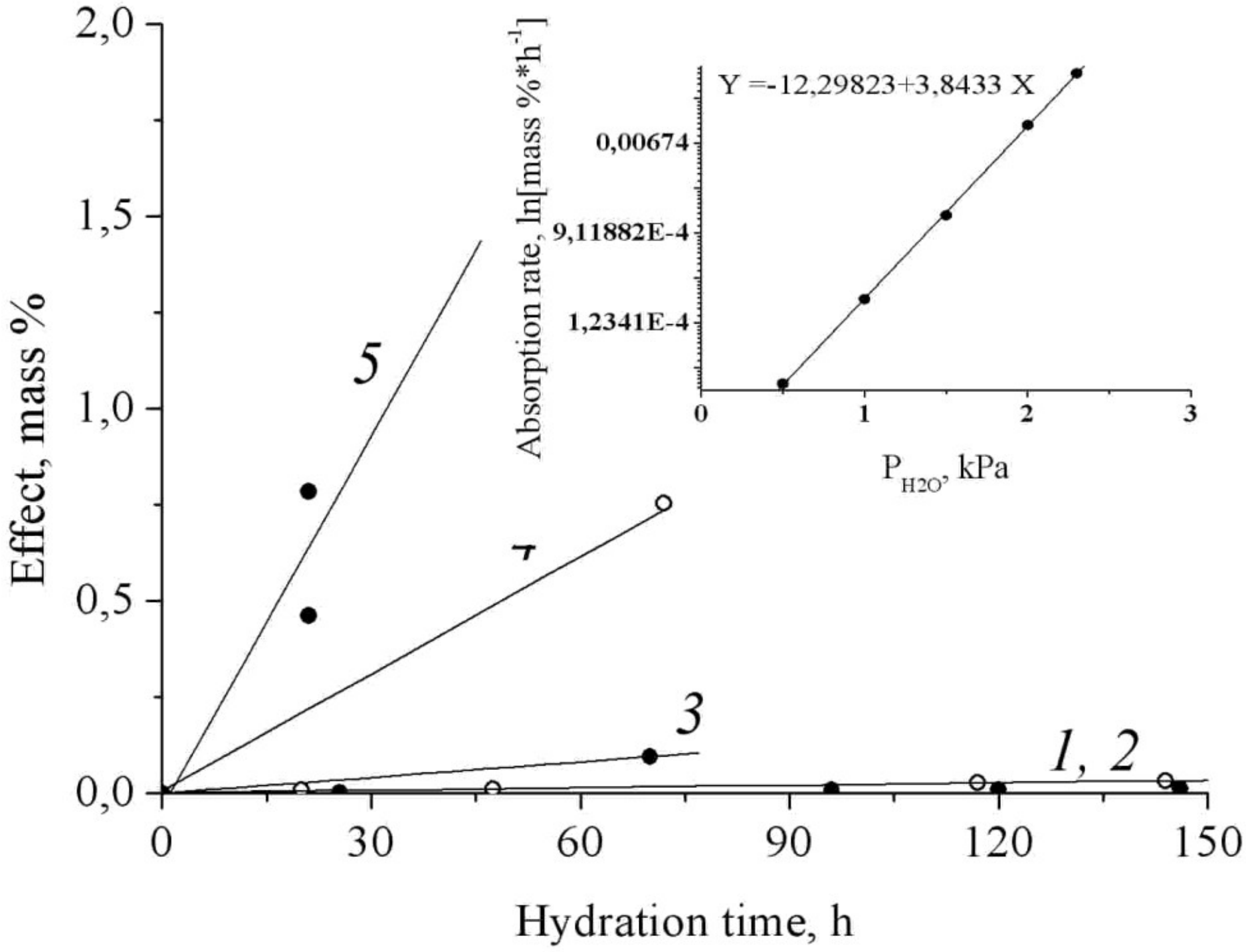

**Fig. 2.** Initial sections of the kinetic absorption curves measured at $p_{H_2O}$ of 0.5 (*1*), 1.0 (*2*), 1.5 (*3*), 2.0 (*4*), and 2.3 kPa (*5*). Inset: linearization of the rate at which gases are absorbed by the $YBa_2Cu_3O_{6.53}$ oxide in the ln($v_{ad}$)–$p_{H_2O}$ coordinates.

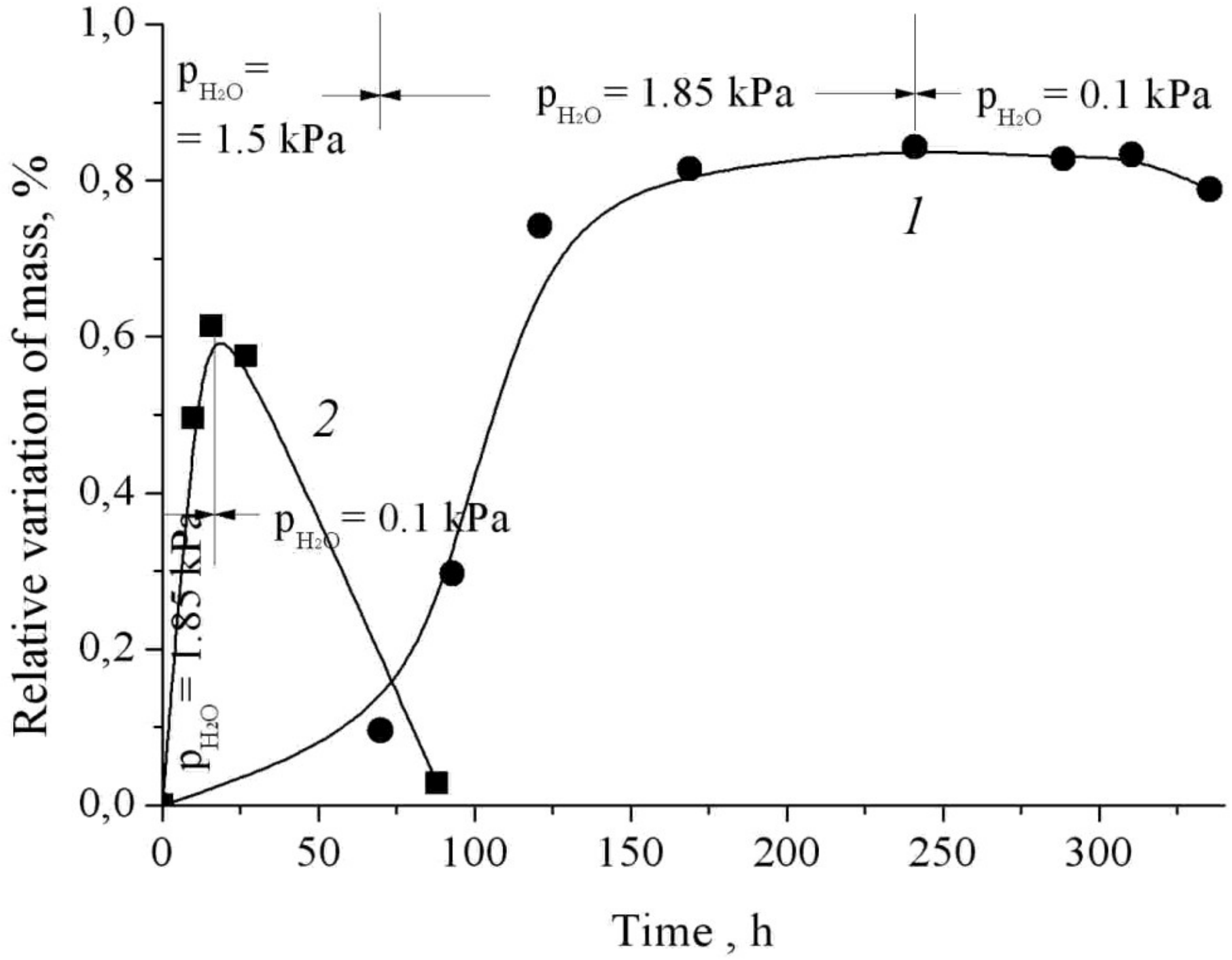

**Fig. 3.** Saturation of $YBa_2Cu_3O_{6.53}$ with gases vs. the drop of $p_{H_2O}$.

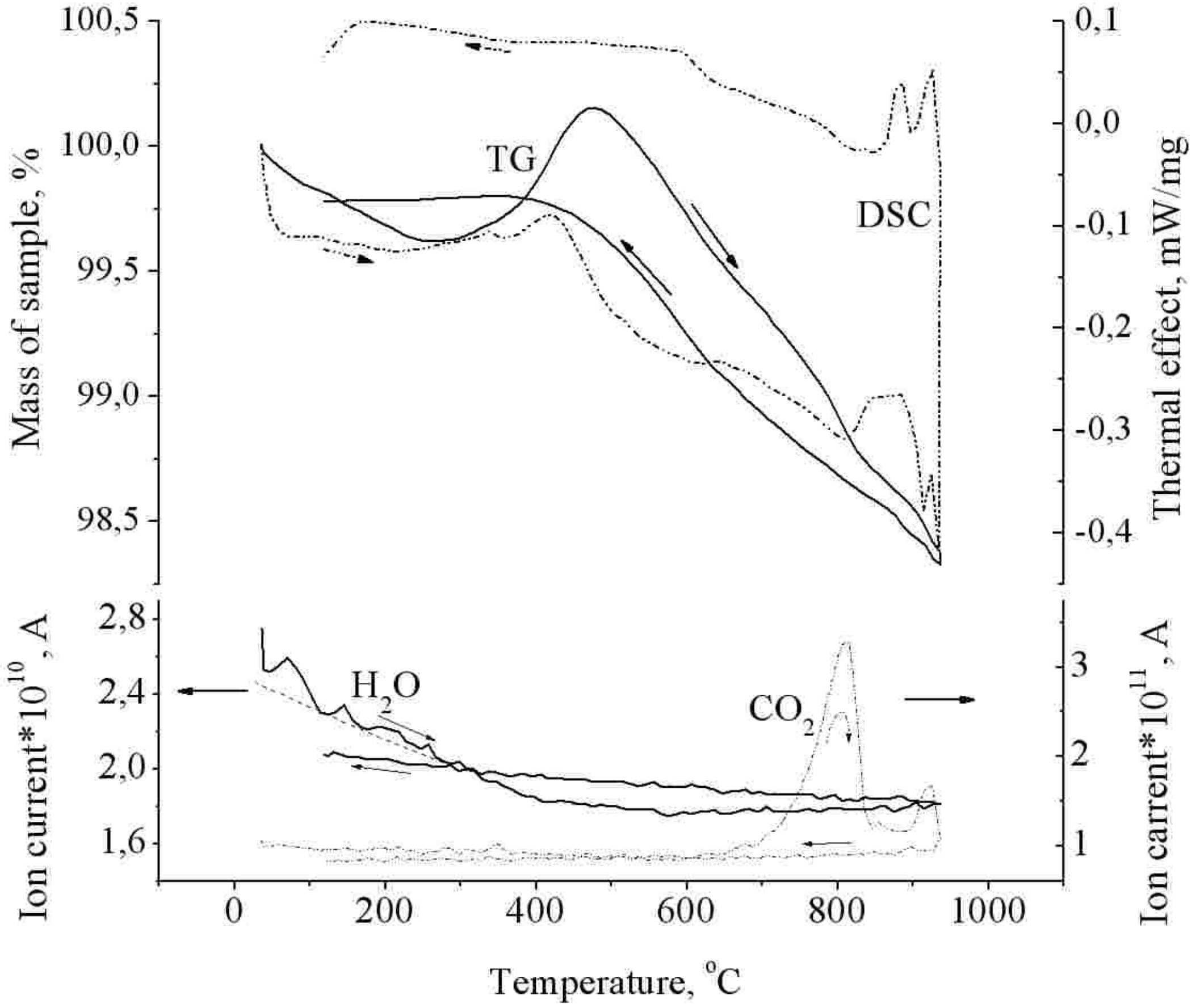

**Fig. 4.** Thermal desorption curves obtained for an $YBa_2Cu_3O_{6.53}$ sample saturated with moisture and $CO_2$ at $p_{H_2O}$ = 2.3 kPa and $p_{CO_2}$ = 0.03 kPa for 21 h. The achieved saturation was 0.78 mass % (the sample corresponds to the first nonzero point in curve 2, Fig. 1).

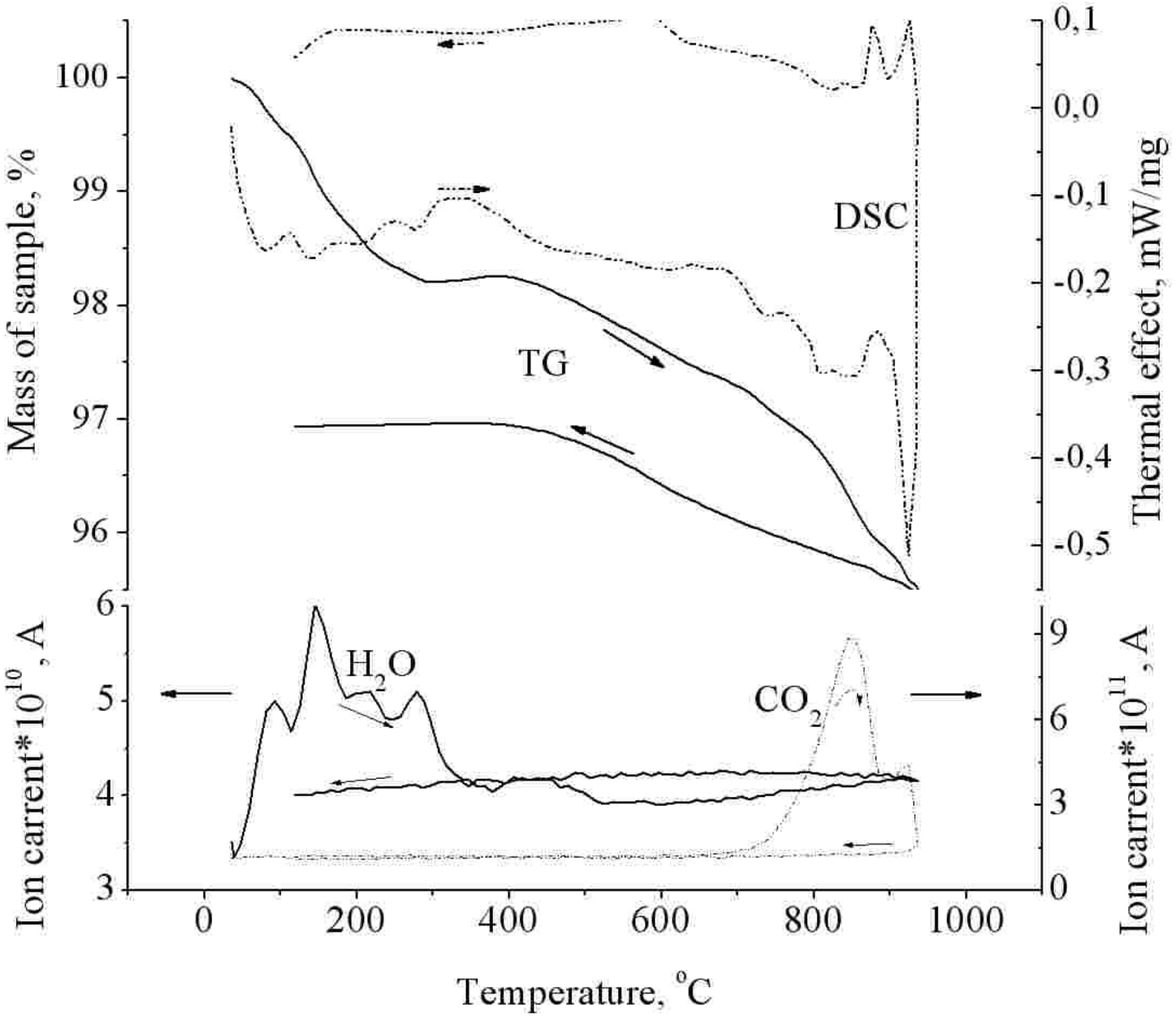

**Fig. 5.** Thermal desorption curves obtained for an $YBa_2Cu_3O_{6.53}$ sample saturated with moisture and $CO_2$ at $p_{H_2O}$ = 2.3 kPa and
$p_{CO_2}$ = 0.03 kPa for 194 h. The achieved saturation was 5.23 mass % (the sample corresponds to the last point in curve 2, Fig. 1).

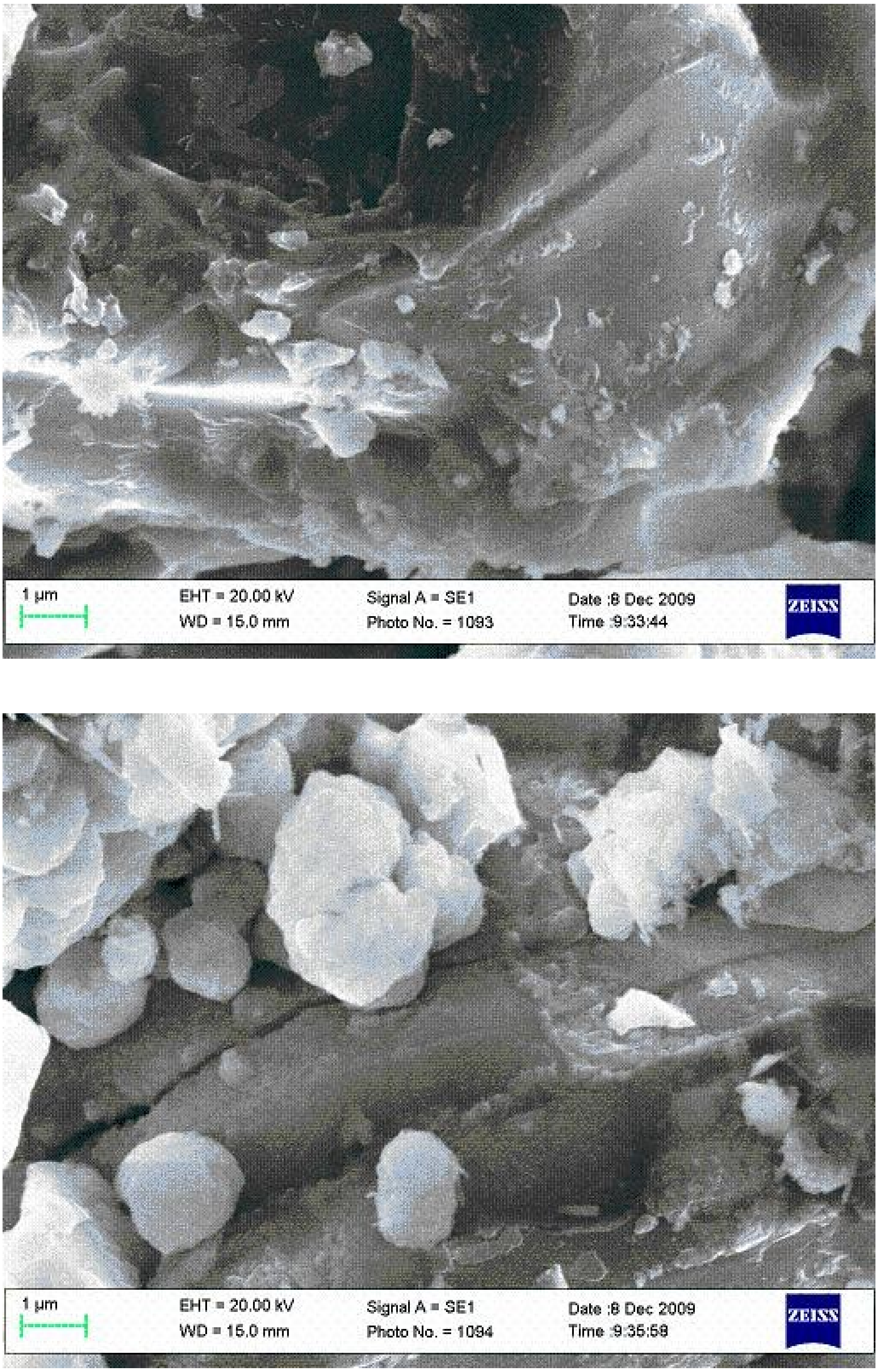

**Fig. 6.** SEM images of the surface of $YBa_2Cu_3O_{6.53}$ particles saturated with gases up to 5 mass % and calcined at 160°C for 1 h; ×20 000.

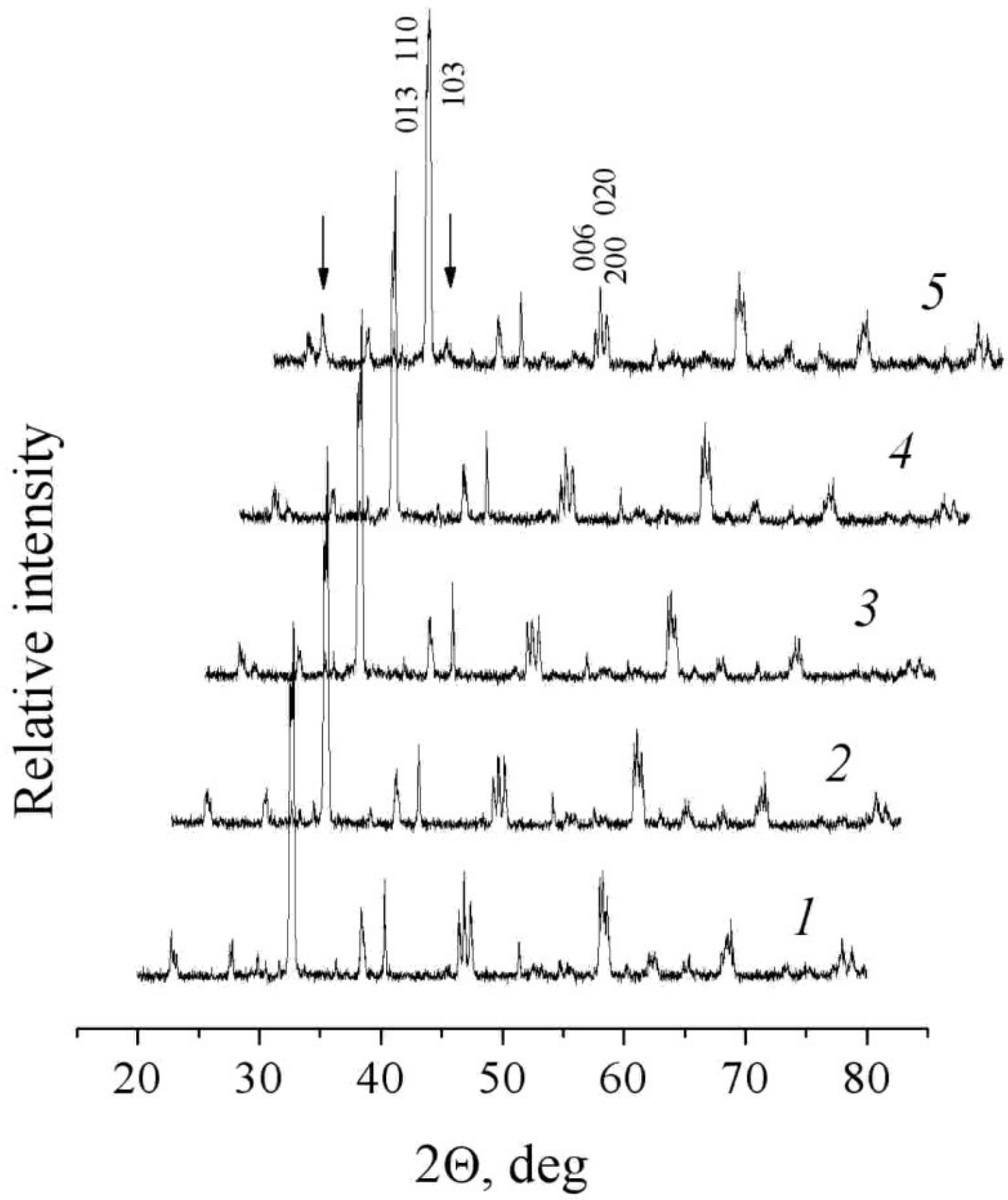

**Fig. 7.** X-ray diffraction patterns of samples taken as $YBa_2Cu_3O_{6.53}$ was consecutively saturated with moisture and $CO_2$ at $p_{H_2O}$ = 2.3 kPa and $p_{CO_2}$ = 0.03 kPa. The saturation level: *1*, 0; *2*, 0.46; *3*, 2.07; *4*, 2.91; *5*, 5.50 mass %.

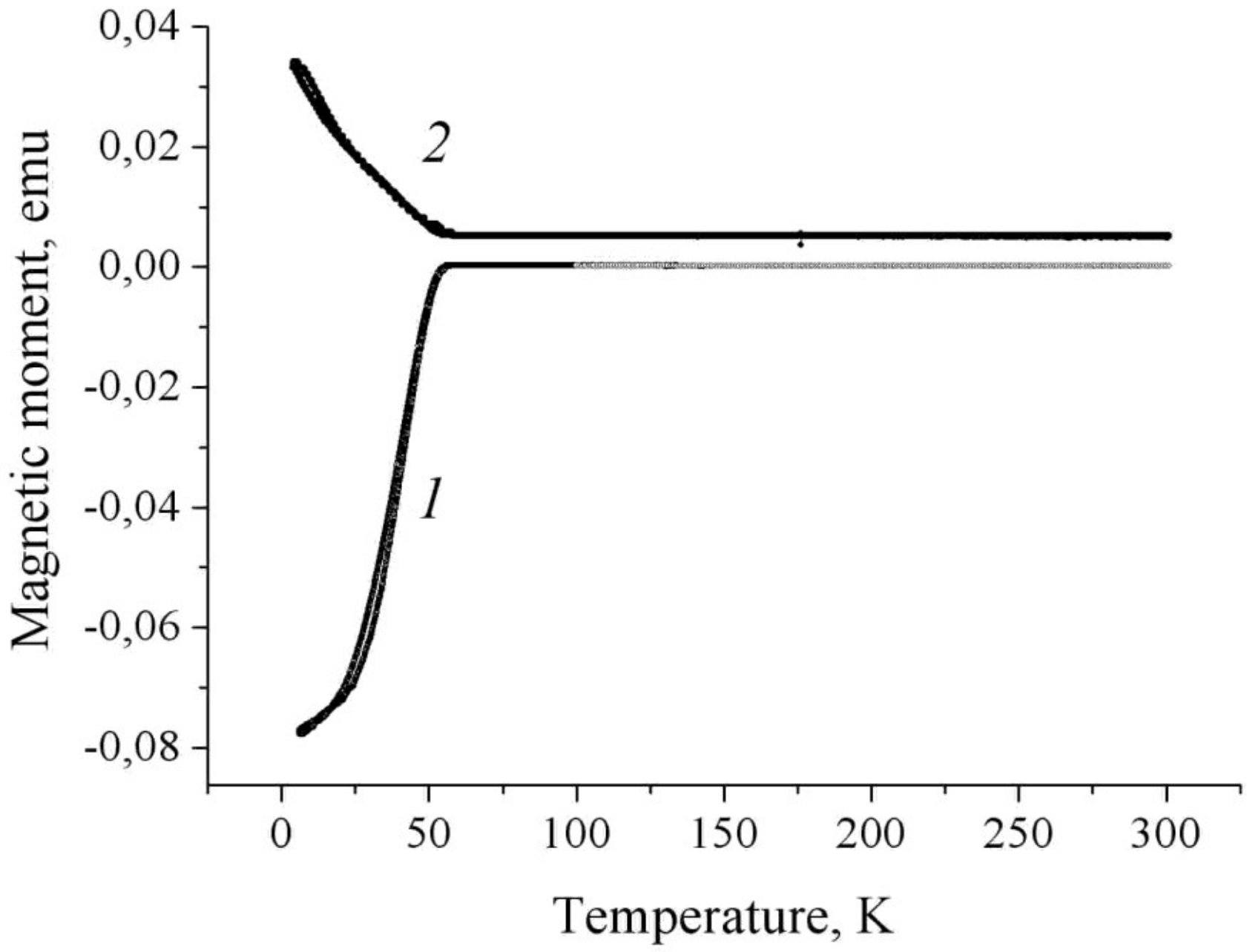

**Fig. 8.** Temperature dependence of the magnetic moment for a "standard" sample of $YBa_2Cu_3O_{6.53}$ (1) and its sample modified as described in [7] (2). The dependence was measured in a field of 0.05 T.

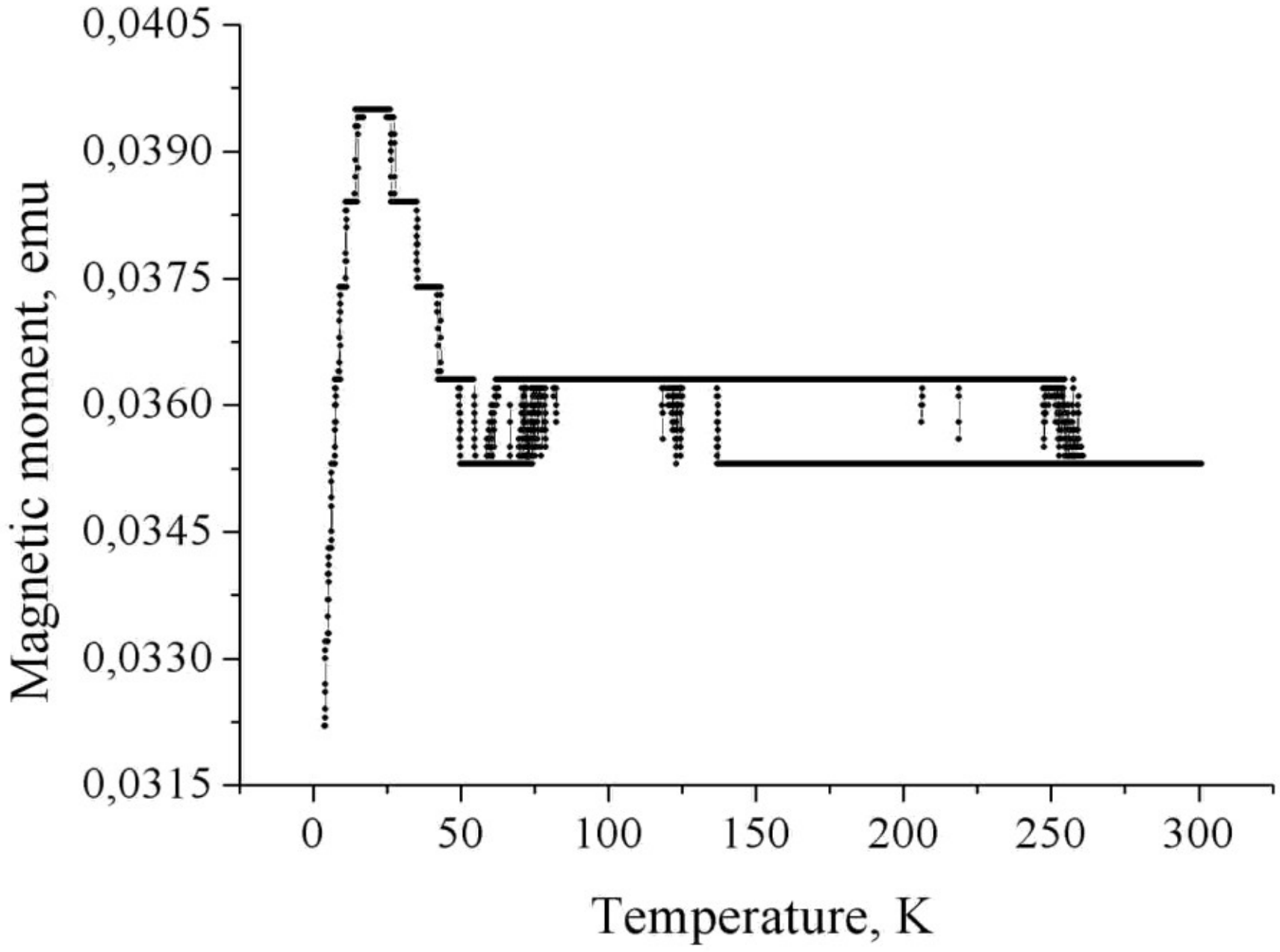

**Fig. 9.** Temperature dependence of the magnetic moment for a modified $YBa_2Cu_3O_{6.53}$ sample as measured in a field of 1 T.